\documentclass[aip,jcp,preprint]{revtex4-1}
\usepackage{graphicx}
\usepackage{dcolumn}
\usepackage{bm}
\usepackage[utf8]{inputenc}
\usepackage{color}
\begin{document}
\title{Spontaneous emergence of autocatalytic information-coding polymers}
\author{Alexei V. Tkachenko}
\email{oleksiyt@bnl.gov}
\affiliation{Center for Functional Nanomaterials, 
Brookhaven National Laboratory, Upton NY 11973} 
\author{Sergei Maslov}
\email{ssmaslov@gmail.com}
\affiliation{Biological, Environmental and Climate 
Sciences Department, Brookhaven National Laboratory, 
Upton NY 11973}
\affiliation{
Department of Bioengineering, 
University of Illinois at Urbana-Champaign,
1270 Digital Computer Laboratory, MC-278
Urbana, IL 61801}
\begin{abstract}
Self-replicating systems based on information-coding polymers are 
of crucial importance in biology. They also recently emerged as 
a paradigm in material design on nano- and micro-scales. We present a general 
theoretical and numerical analysis of the problem of spontaneous 
emergence of autocatalysis for heteropolymers capable of 
template-assisted ligation driven by cyclic changes in the environment. 
Our central result is the existence of the first order transition 
between the regime dominated by free monomers and that with a 
self-sustaining population of sufficiently long chains. 
We provide a simple, mathematically tractable model supported by 
numerical simulations, which predicts the distribution of 
chain lengths and the onset of autocatalysis 
in terms of the overall monomer concentration and two 
fundamental rate constants.  Another key result of our study 
is the emergence of the kinetically-limited optimal 
overlap length between a template and each of its two substrates. 
The template-assisted ligation allows for heritable transmission 
of the information encoded in chain sequences 
thus opening up the possibility of long-term memory and evolvability 
in such systems.
\end{abstract}
\pacs{}

\maketitle

\section{Introduction}

    Life as we know it today depends on self-replication of information-coding polymers. 
Their emergence from non-living matter is one of the greatest 
mysteries of fundamental science. In addition, the design of artificial self-replicating nano- and micro-scale systems is an exciting field with potential 
engineering applications \cite{Chaikin,Brenner}. The central challenge in both of 
these fields is to come up with 
a simple physically-realizable system obeying laws of thermodynamics,
yet ultimately capable of Darwinian evolution when exposed 
to non-equilibrium driving forces. 
Chemical networks of molecules engaged in mutual catalysis 
is a popular candidate for such a system\cite{Eigen,Dyson,Kauffman,Jain}. 
Information-coding heteropolymers 
In fact, the most successful experimental realization of an autonomous 
self-replicating system involves a set of mutually catalyzing 
RNA-based enzymes (ribozymes)\cite{Szostak_1989}
that show evolution-like behavior\cite{Joyce}.
This is viewed as a major evidence for RNA-world hypothesis 
(see e.g. Refs.  \onlinecite{Gilbert,Orgel,Joyce_Review}. )

The ribozyme activity requires relatively 
long polymers made of hundreds of nucleotides with 
carefully designed sequences. 
Polymers of sufficient length can be generated e.g. by traditional 
reversible step-growth polymerization that
combines random concatenation and fragmentation of 
polymer chains. Furthermore, the 
polymer length in this type of process 
can be drastically increased 
in non-equilibrium settings such as temperature 
gradients \cite{Braun}. However, even when 
long chains are formed, the probability 
of the spontaneous emergence 
of a sequence with an enzymatic activity 
remains vanishingly small,  due to the 
exponentially large number of possible sequences. 

Thus there is a strong need for a mechanism 
that combines the emergence of long chains with 
dramatic reduction of informational entropy of 
the sequence population. A promising candidate 
for such mechanism is provided by template-assisted ligation. 
In this process pairs of polymers are brought together 
by hybridization with a complementary 
polymer chain serving as the template and 
eventually ligated to form a longer chain. Unlike the 
non-templated reversible step-growth polymerization 
used in Ref. \onlinecite{Braun}, 
this mechanism naturally involves the information transmission
from a template to the newly-ligated chain, thus opening an exciting 
possibility of long-term memory and evolvability.
An early model involving template-assisted polymerization
was proposed by P. W. Anderson and 
colleagues\cite{Anderson_Stein,Anderson}. It also has been 
subject of more recent experimental and theoretical 
studies\cite{Szostak_1996,Derr}. In particular, in 
Ref. \onlinecite{Derr} it has been demonstrated that, for a specific choice of 
parameters, a combination of non-template and template-assisted ligation
can lead to the emergence of long (around 100 monomers) 
oligonucleotides.

In this work we carried out the theoretical and 
numerical analysis of a generic system in which 
the polymerization is driven solely by template-assisted 
ligation. Unlike in the models with significant contribution of 
non-templated concatenation, the emergence of long chains in our 
system represents a non-trivial chicken-or-egg problem.
Indeed, the formation of long chains depends on the 
presence of other chains serving as templates. 


In our model the ``primordial soup'' of monomers 
is driven out of equilibrium by cyclic changes in 
physical conditions such as temperature, 
salt concentration, pH, etc. (see Fig. 1ab).
Polymerization occurs during the ``night'' phase
of each cycle when the existing heteropolymers serve as templates for 
formation of progressively longer chains. 
During the ``day'' phase of each cycle 
all multi-chain structures separate and the system returns 
to the state of dispersed individual polymers.

We consider a general case of information-coding heteropolymers 
composed of $z$ types of monomers 
capable of making $z/2$  mutually complementary pairs. 
For example, RNA is made of $z=4$ monomers forming $2$ complementary 
pairs $A-U$ and $C-G$ responsible for double-stranded RNA 
structure. Similarly, we assume that hybridization 
between complementary segments of our 
generalized polymers results in formation of a 
double-stranded structure. During the night phase 
of each cycle chains form a variety of hybridized complexes.
The ligation takes place in a special type of 
such complexes shown in Fig. 1b. 
The end groups of two 
``substrate'' chains  $S_1$ and $S_2$
are positioned next to each other
by the virtue of hybridization with the 
third, ``template'' polymer $T$.
Once the substrates are properly positioned, the new 
covalent bond joining them together is formed at a constant rate. 
We further assume 
that each of the intra-polymer bonds can spontaneously 
break at a constant rate making the overall fragmentation 
rate of a chain proportional to its length.
\begin{figure}
\centering
\includegraphics[width=1\columnwidth]{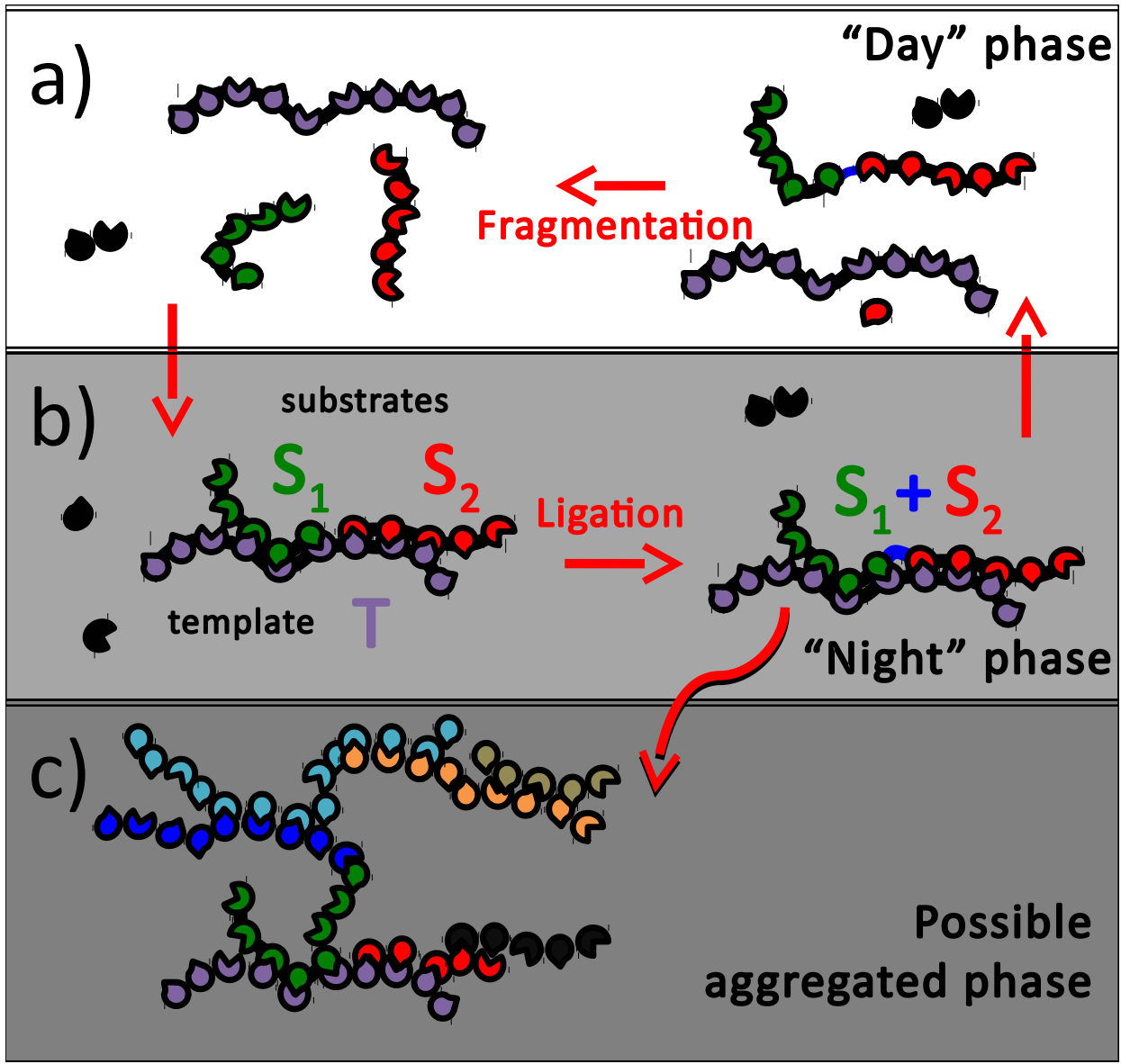}
\caption{\it {\bf 
The schematic representation of fundamental processes 
in our system. }a) The "day" phase during which 
all hybridized complexes between heteropolymers dissociate and 
ligation completely stops, while fragmentation continues in all 
phases of the cycle.
b) The "night" phase during which some 
polymer chains hybridize and then undergo
template-assisted ligation.
The ends of substrates $S_1$ (green) and $S_2$ (red)
hybridized with a template $T$ (purple) are ligated at a constant rate
with the newly formed bond shown in blue.
c) If the "night" phase is sufficiently long heteropolymers enter 
the aggregation regime in which ligation effectively stops.
}
\label{fig1}
\end{figure}
If one was to leave a mixture of polymers  
in the night phase long enough, 
hybridization of multiple chains 
would result in the formation of a gel-like
aggregate shown in Fig. 1c, 
effectively stopping ligation. 
During the day phase of the cycle (Fig. 1a) 
all structures of hybridized polymers 
dissociate while keeping their stronger internal bonds intact. 
Thus the day phase plays the role of the ``reset'' 
returning the system to a mixture of free polymers 
ready for the next night phase.

One of the major assumptions used in our study is the Random Sequence Approximation (RSA)
according to which each monomer in every chain can be of any type
with equal probability $1/z$.  On the one hand, the RSA greatly simplifies 
the problem and allows us to get a concise analytical solution. On the other hand, 
in order to understand the transmission of sequence-encoded information and 
the long-term memory in our system this approximation need to 
be relaxed in future studies. 

\section{Results}

\subsection{Optimal overlap length $k_0$}

In general the 
interaction strength between any two chain segments
increases with the overlap length $k$ of the region over which they are 
complementary to each other. Here we assume a simple linear 
relationship in which the binding free energy 
is given by $\Delta G_0+k \cdot \Delta G$, where $\Delta G$ 
is the (negative) binding free energy 
between two complementary monomers, while $\Delta G_0$ is the
initiation free energy. 



The equilibrium hybridization probability 
emerges out of the competition between two opposing
kinetic processes of association and dissociation.
On the one hand, the association rate exponentially decreases
with the overlap length $k$ since the probability of finding 
a pair of polymers with complementary 
sequences of length $k$, is proportional to $1/z^k$. 
On the other hand, the dissociation rate between 
a substrate and its template also exponentially 
decreases with $k$ as $\exp(-k \cdot \Delta G/k_{B}T)$ due to 
greater thermodynamic stability of longer complementary 
duplexes.  The net result is that 
the hybridization probability is proportional to 
$\exp(k \cdot \epsilon)$, where 
\begin{equation}
\epsilon=-\Delta G/k_{B}T-\log(z)
\label{eq_epsilon}
\end{equation}
is the effective parameter combining 
thermodynamic and combinatorial factors.
Template-assisted ligation happens at appreciable rates 
only for $\epsilon>0$, i.e. when
$\Delta G<-k_{B}T\log(z)$. 
For a finite time window $t$ only the duplexes with 
short overlaps  will reach this 
equilibrium. Duplexes with longer overlaps
have lifetimes much longer than $t$. Thus for them 
the hybridization probability is limited by the 
association rate alone $\sim 1/z^k$.
Therefore, the overall hybridization probability as a function of $k$
is strongly peaked (see Fig. 2 and Appendix A for details). 
As time $t$ increases, this peak slowly (logarithmically in $t$) 
shifts  towards larger values of $k$
with its final value $k_0$ set by 
either the end of the night phase or 
(in case of long nights) by 
the onset of the aggregation 
regime (Fig. 1c). 
\begin{figure}
\centering
\includegraphics[width=1\columnwidth]{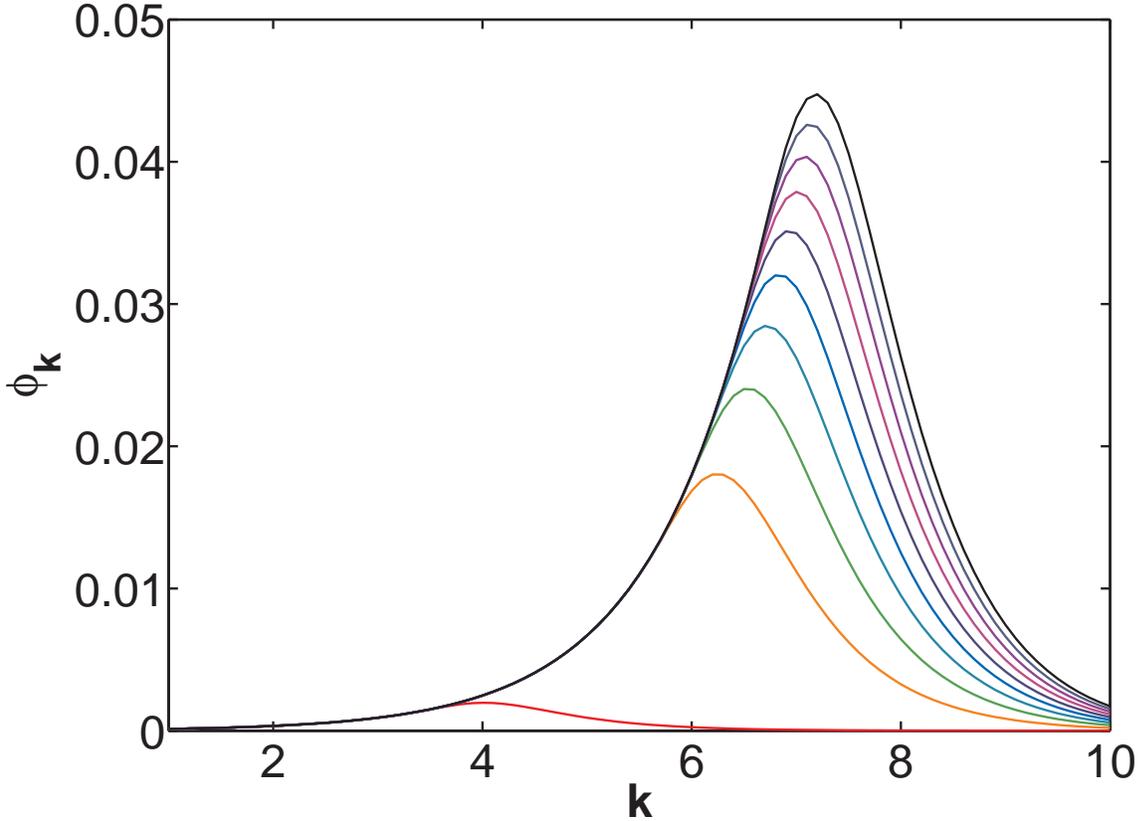}
\caption{\it
{\bf Time evolution of the hybridization probability.}
The probability that a segment of length $k$ 
is hybridized to its complementary partner (Eq. (\ref{S8}) in Appendix A)
is strongly peaked at $k=k_0 \sim \log t$ (see Eq. (\ref{S9}) in Appendix A). 
Different colors from red to violet correspond to 
linearly increasing times $t$ 
since the beginning of the night phase of the cycle. 
}
\label{fig2}
\end{figure}

\subsection{Major parameters of the model}
In what follows we focus on slow 
dynamical processes taking place over multiple day/night cycles.
The main input parameter from the intra-night kinetics
to the multi-cycle dynamics is 
the hybridization cutoff length $k_0$ discussed above. 
The multi-cycle dynamics can be described in terms of time-averaged 
ligation and fragmentation rates, $\lambda$ and $\beta$, respectively. 
We define $\lambda$ as the rate of bond 
formation 
provided that the ends of the two substrates are already properly 
positioned next to each other due to their hybridization 
with the template.
We further assume that 
the characteristic fragmentation time $1/\beta$ is much longer than the duration of 
the day-night cycle ensuring the separation between short and long timescales in the problem.
Both $\lambda$ and $\beta$ are averaged over the duration of the 
day-night cycle with the understanding that fragmentation happens continuously 
throughout the cycle (possibly with different day and night rates), 
while the ligation only occurs during the night phase. Thus $\lambda$ 
implicitly depends on relative durations of night and day phases.

Let $C$ be the overall monomer concentration including both free monomers 
and those bound in all chains. 
In the case of random sequence composition, 
the population of heteropolymers is fully characterized by their length distribution $f_l$, 
defined in such a way that $C \cdot f_l$ is the concentration of all
polymers of length $l$. By this definition $f_l$ is subject 
to the normalization condition $\sum_{l=1}^{\infty} l f_l=1$.
The fraction of polymers with a 
specific sequence is then given by $z^{-l}\cdot f_l$.

\subsection{Detailed balance ansatz}
For template-assisted ligation the effective two-polymer merger rate $\mu$ 
is given by the ligation rate 
$\lambda$ multiplied by the probability 
of hybridization of a template $T$ with two 
substrates $S_1$ and $S_2$ bring them into 
end-to-end configuration shown in Fig. 1b. 
The major step in constructing an approximate 
analytical solution of the problem 
is the {\it assumption of a detailed balance} 
between template-assisted ligation and fragmentation 
in the steady state of the system:
\begin{equation}
\beta f_{l+m}=\mu f_l \cdot f_m \qquad .
\label{eq1}
\end{equation}
Here, the left-hand side describes the rate at which 
a chain of length $l+m$ breaks into two pieces of lengths
$l$ and $m$ correspondingly. Conversely, the right-hand side is the 
effective merger rate (hybridization + ligation) at which 
polymers of lengths $l$ and $m$ are joined to form a longer chain of
length $l+m$. Note that according to this description 
the rate at which a polymer breaks into arbitrary two 
pieces is proportional to its length or rather 
its number of intra-polymer bonds.

The detailed balance approximation is not a priory justified 
in driven, non-equilibrium systems such as ours.
However, for chains longer than the optimal overlap length $k_0$ 
the probability of hybridization and thus 
the effective merger rate $\mu$ saturate
(see Methods for derivation and details) . 
Once both $\mu$ and $\beta$ are independent of polymers' lengths, 
our system becomes mathematically 
equivalent to the well known reversible step-like polymerization process
for which the detailed balance approximation hold true by the virtue
of laws of equilibrium thermodynamics. 
In spite of this superficial similarity,  
our system remains intrinsically non-equilibrium since the effective
merger rate $\mu$ depends on hybridization between templates and substrates 
cycled through day and night phases as shown in Fig. 1ab.
In addition, the Eq. (\ref{eq1}) is expected to break down for chains shorter 
than $k_0$. 

To validate our mathematical 
insights, the analytic solution shown below was followed by 
numerical simulations of the system carried out {\it without 
the detailed balance approximation}. 
The agreement between our analytical and numerical results
for polymers longer than $k_0$ 
confirms the validity of our approach. 

The Eq. (\ref{eq1}) 
is satisfied by the exponential length distribution:
\begin{equation}
f_l=(\beta/\mu)\exp(-l/\bar{L}) \quad ,
\label{eq_fl}
\end{equation}
where the characteristic chain length, $\bar{L}$, 
is determined by the normalization condition 
$\sum_{l=1}^{\infty} l f_l=1$ 
or $(\beta/\mu) \bar{L}^2 =1$.
This result was obtained by replacing the discrete
sum with the integral, which works in the limit 
$\bar{L} \gg 1$ (see Eq. \ref{SX} in 
Appendix B for the exact formula in which 
this approximation is relaxed).
Hence, the characteristic chain length 
in the steady state exponential distribution is given by
\begin{equation}
\bar{L}=\sqrt{\frac{\mu}{\beta}}
\label{eq_Lbar}
\end{equation}

\subsection{Onset of autocatalysis}
$\mu$ is an effective two-polymer merger rate 
proportional to the probability of finding two terminal 
ends attached to a template followed by ligation. 
This probability depends 
on (a) the overall concentration $C$ and the length 
distribution of potential templates (b) the strength 
and kinetics of interactions between the complementary segments 
on a template and its two substrates. 

For short overlaps $k \leq k_0$ the hybridization probability 
follows the equilibrium formula:
$\sim \exp(k \cdot \epsilon)$. This increase is followed by 
an abrupt drop for $k>k_0$ (see Fig. 2). 
By neglecting the contribution of overlap lengths longer than $k_0$ one gets
\begin{eqnarray}
\mu&=&\lambda \left(\frac{C}{C_0}\right)^2
\sum_{k_1=1}^{k_0}\exp(k_1 \cdot \epsilon) \sum_{k_2=1}^{k_0}\exp(k_2 \cdot \epsilon) \cdot \nonumber \\
&\cdot &\sum_{l=k_1+k_1}^{\infty}(l-k_1-k_2+1)f_l \quad .
\label{eq_lambda} 
\end{eqnarray}
Here $\lambda$ is the bare ligation rate, 
$k_1$ and $k_2$ are the overlap lengths between the template
and each of the two substrates. 
We also introduced the reference concentration 
$C_0=\exp[-\Delta G_0/k_{B}T]$ (in molar)
absorbing the initiation free energy. 
The term $(C/C_0)^2$ reflects the fact that the template-assisted
ligation is a three-body interaction involving two substrates and one 
template. The last sum in the r.h.s. of the Eq. (\ref{eq_lambda}) 
is equal to the probability of finding a template region of length $k_1+k_2$
within a longer heteropolymer. It takes into account that a chain 
of length $l \geq k_1+k_2$ has $l-k_1-k_2+1$ sub-sequences of length $k_1+k_2$.
Requirements of sequence complementarity between the template and each of 
two substrates  were absorbed into the definition of $\epsilon$ within the RSA. 

Substituting the exponential distribution $f_l$ given by the 
Eq. (\ref{eq_fl}), performing the triple summation in  Eq. 
(\ref{eq_lambda}), and neglecting the terms $\sim 1/\bar{L}$ 
(but not $\sim k_0/\bar{L}$) within the exponents approximately gives
$\mu=\lambda (C/C_0)^2\exp(2k_0 \cdot (\epsilon -1/\bar{L}) )/\left[1-\exp(-\epsilon)\right]^2$. 
Substituting this expression into the Eq. (\ref{eq_Lbar}) results in the 
self-consistency equation for $\bar{L}$:
\begin{equation}
\bar{L}\exp\left(\frac{k_0}{\bar{L}}\right)=\frac{C}{C_0} \cdot \sqrt{\frac{\lambda}{\beta}} 
\cdot \frac{\exp\left(k_0 \epsilon \right)}{1-\exp(-\epsilon)} \quad ,
\label{eq_sc_SI}
\end{equation}
(see the Eq. (\ref{eq_sc}) in the Appendix B for a more precise expression
derived without the large $\bar{L}$ approximation).
The l.h.s. of this equation reaches its minimal value of $e \cdot k_0$ at 
$\bar{L}=k_0$. As a result, the equation has solutions only for 
concentrations $C$ above a certain threshold value given by
\begin{equation}
C_{down}=k_0C_0\sqrt{\frac{\beta}{\lambda}}\exp(1-k_0 \epsilon)\cdot (1-\exp(-\epsilon))
\label{eq_c_down}
\end{equation}
For $C$ significantly larger than this threshold, one can neglect the exponential term 
in the l.h.s. of the Eq. (\ref{eq_sc_SI}) so that the characteristic 
polymer length $\bar{L}$ linearly increases with the concentration as 
\begin{equation}
\bar{L}=\frac{C}{C_0} \cdot \sqrt{\frac{\lambda}{\beta}} \cdot \frac{\exp(k_0 \epsilon)}
{1-\exp(-\epsilon)} \quad .
\label{eq_lbar_vs_C}
\end{equation}
For monomer concentrations $C$ below the threshold we don't expect 
long heteropolymers to form. This suggests a first-order 
transition between the regimes dominated by free monomers 
and that with a self-sustaining population of long heteropolymeric chains.

To verify and refine our predictions we approach this transition 
from below, starting with the state dominated by monomers i.e.
$f_1 \simeq 1$.
We explore the stability of the monomer mixture with respect to 
formation of dimers. 
In this limit, the dimer fraction $f_2$ obeys the 
following kinetic equation:
\begin{equation}
\frac{df_2}{dt}=-\beta f_2+\lambda \left(\frac{C}{C_0}\right)^2 \exp(2\epsilon) f_1^2f_2 \qquad ,
\label{eq_dimer}
\end{equation}
where the second term in the r.h.s. reflects the fact that a dimer can be formed out of 
two monomers and this process needs to be catalyzed by a complementary dimer.
The critical concentration $C_{up}$ above which 
dimers would exponentially self-amplify is given by 
\begin{equation}
C_{up}=C_0 \sqrt{\frac{\beta}{\lambda}}\exp(-\epsilon) \qquad .
\label{eq_c_up}
\end{equation}
Thus we confirm the existence of an instability in a mixture of monomers 
with respect to template-assisted formation of longer chains.
Note that, as expected for a first-order phase transition, 
the instability threshold $C_{up}$ (Eq. (\ref{eq_c_up})) approached from below 
exceeds the instability threshold $C_{down}$ (Eq. (\ref{eq_c_down}))
approached from above. Thus, as expected for a first-order phase transition, 
the system will be hysteretic for $C_{down}<C<C_{up}$.

\subsection{Numerical results}
To check our calculations we carried out the detailed 
numerical simulations of our system. Specifically, 
we numerically solved a system of coupled kinetic 
equations describing the template-assisted 
ligation and fragmentation processes and 
calculated the steady state distribution $f_l$:
\begin{eqnarray}
\frac{1}{2\beta}\dot{f}_{n} &=& -\left[  \frac{n}{2}+\Gamma^{2}\sum\limits_{m}%
\mu_{n,m}f_{m}\right]  f_{n}+\sum\limits_{m>n} f_{m}+\nonumber \\
&&+\Gamma^{2}\sum\limits_{m<n}\frac{\left(
1+\delta_{n-m,m}\right)  }{2}\mu_{m,n-m}f_{m}f_{n-m} \qquad . 
\label{kin2_detailed}%
\end{eqnarray}
Here $\Gamma$ is the dimensionless control parameter of 
the model proportional to the monomer concentration:
\begin{equation}
\Gamma=\left(\frac{C}{C_{0}}\right)\sqrt{\frac{\lambda}{\beta}} \label{nu}%
\end{equation}
and $\mu_{nm}$ is the merger matrix, which itself 
linearly depends on the distribution $f_l$ as described 
in Eq. (\ref{S17}) and (\ref{S1}) in Appendices A and C.  
Note, that these simulations (unlike our analytical theory) 
allow for overlap length dependence of merger rates
do not use the detailed balance ansatz.

The results of these numerical simulations are in excellent agreement 
with our analytical calculations. For high enough concentrations $C$ 
the length distribution $f_l$ has a long exponential tail
covering the region $l>k_0$. Chains of length shorter than $k_0$, which
do not obey the detailed balance, exhibit a much faster decay 
as a function of $l$ (see Fig. 3). 
\begin{figure}
\centering
\includegraphics[width=1\columnwidth]{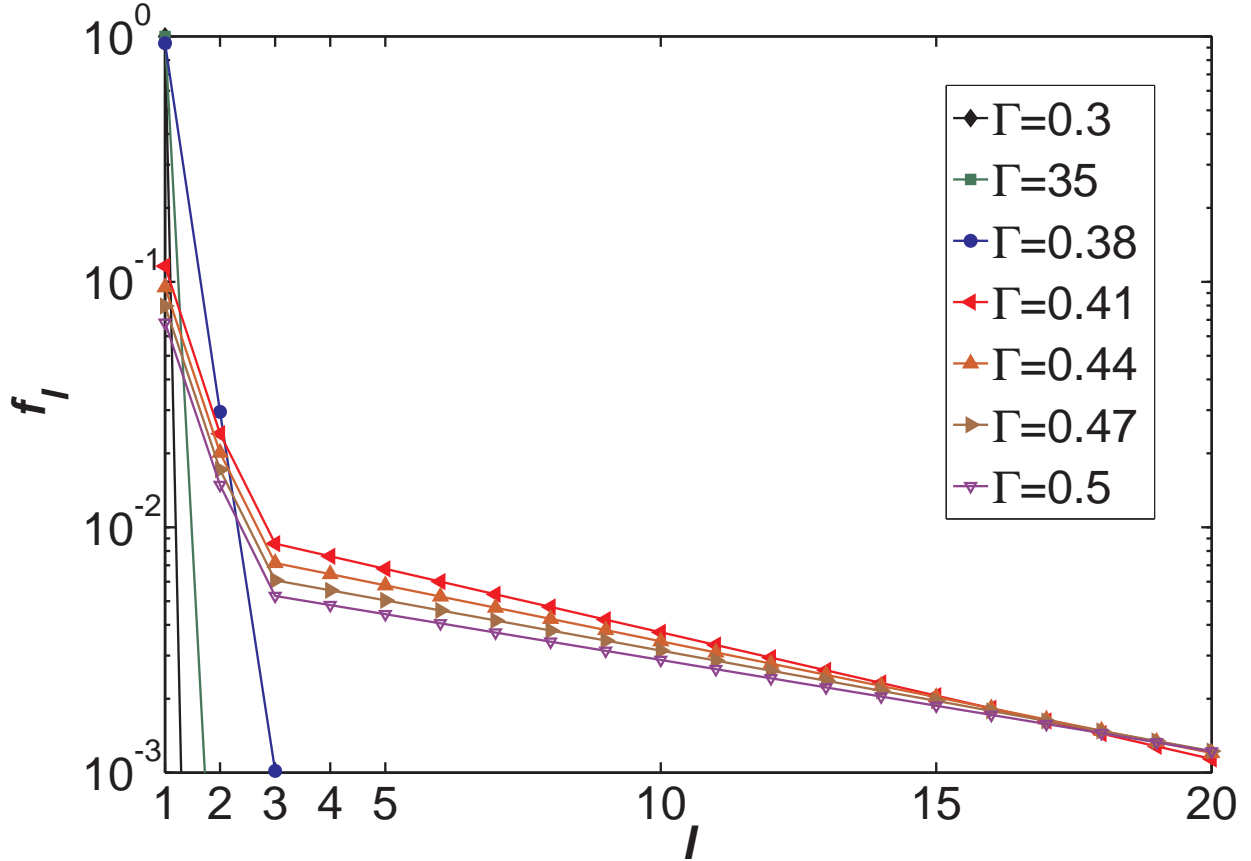}
\caption{\it
{\bf Chain length distributions.}
A set of chain length distributions $f_l$ 
plotted for different values of the control parameter 
$\Gamma=\frac{C}{C_0}\sqrt{\frac{\lambda}{\beta}}$ as 
found by numerical simulations with $k_0=3$ and $\epsilon=1$. 
Distributions in the autocatalytic regime are characterized by 
long exponentially distributed tails for chains with
$l>k_0$. Note a sharp transition between monomer-dominated 
and autocatalytic regimes.} 
\label{fig3}
\end{figure}

Our simulations also confirmed the existence of a first-order 
transition to a regime dominated by monomers as concentration $C$ 
was reduced (the red line in Fig. 4). 
The decay length $\bar{L}$ of the exponential tail of $f_l$ 
for $l \geq k_0$ plays the role of the order parameter in this transition.
When plotted as a function of concentration $C$ in Fig. 4,
it exhibits sharp discontinuities and hysteretic behavior. 
Our analytical results given by the 
Eq. (\ref{eq_sc}) (black dashed line in Fig. 4) 
are in a good agreement with our numerical simulations. 
%
\begin{figure}
\centering
\includegraphics[width=1\columnwidth]{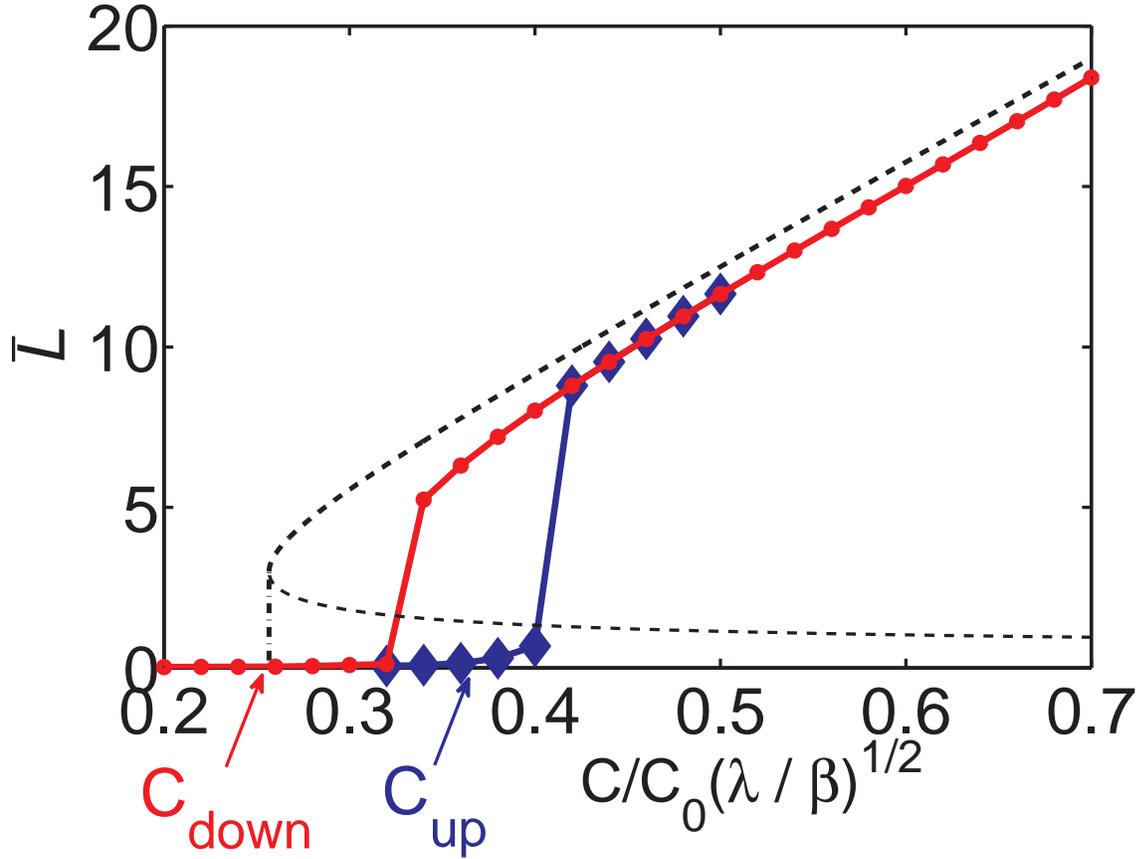}
\caption{\it
{\bf A hysteretic first order transition between the monomer-dominated 
and autocatalytic regimes.} Different lines/symbols show 
the characteristic length $\bar{L}$ in our
numerical simulations with $k_0=3$ for increasing (diamonds), and 
decreasing (circles) concentration $C$, correspondingly.
The dashed line is the prediction of our simplified model given by 
the Eq. (\ref{eq_sc}). Arrows indicate $C_{up}$ and $C_{down}$
given by Eqs. (\ref{eq_c_up}) and (\ref{eq_c_down}) correspondingly.
}
\label{fig4}
\end{figure}
The transitions from monomers to long-chained polymers and back 
in our numerical simulations occur at concentrations somewhat higher than 
their theoretically predicted values $C_{up}$ 
(Eq. \ref{eq_c_up}) and $C_{down}$ (Eq. \ref{eq_c_down}) 
marked in Fig. 4 by the blue and red arrows respectively. 

\subsection{Long-night limit}
Our model assumes cyclic changes between "day" and "night" 
phases. In the beginning of each night phase all polymers 
are unhybridized, but as time progresses they start forming duplexes
of progressively longer lengths. The probability of 
finding any given segment in a duplex remains low at the early 
stage of this process. However, if the duration of the 
night phase is long enough, there would be a time point at 
which individual polymers would on average have 
around one hybridized partner. 
Note that a single polymer may simultaneously have more than 
one hybridized partner as long as the duplexes
with different partners do not overlap with each other. 
Around this time most polymers in our pool would become immobilized 
in a gel-like structure schematically depicted in Fig. 1c.
At this point the formation of new 
hybridized complexes effectively stops and the value of 
$k_0$ stops growing. An indirect experimental evidence 
for such aggregation phase was recently reported by Bellini 
{\it et al.}\cite{Bellini}.

According to our results, 
the characteristic chain length $\bar{L}$ 
given by Eq. (\ref{eq_lbar_vs_C}) exponentially increases 
with $k_0$. 
In the presence of aggregation this growth 
is eventually arrested. The upper bound 
on $\bar{L}$ reached in this case 
can be determined self-consistently by requiring 
that individual polymers on average have 
around one hybridized partner. 
A chain of length $\bar{L} \gg k_0$ 
contains $\bar{L}-k_0+1 \simeq \bar{L}$ 
segments of length $k_0$. 
The probability of each of these segments 
to be hybridized at any particular
time is $(C/C_0) \cdot \exp(k_0 \cdot \epsilon)$. 
Thus the transition to the aggregated state is expected when 
\begin{equation}
\bar{L}\frac{C}{C_0} \exp(k_0\epsilon) \simeq 1 \quad .
\label{eq_gel}
\end{equation}
Combining this expression with 
the Eq. (\ref{eq_lbar_vs_C}) 
and ignoring the factors of order of 1 
one gets the upper bound
$\bar{L}_{max}$ on the characteristic 
polymer length that could, in principle, be reached 
by increasing the duration of the night phase:
\begin{equation}
\bar{L}_{max} \simeq 
\left(\frac{\lambda}{\beta}\right)^{\frac{1}{4}} 
\qquad .
\label{eq_Lmax}
\end{equation}

\section{Discussion}
To summarize, above we considered a general 
case of random heteropolymers 
capable of template-assisted ligation. 
As such our model is applicable to 
both nucleic acids at the dawn of life as well as 
to 
artificial 
self-replicating nano- or micro- 
structures\cite{Chaikin,Brenner}. 
The major conclusions of our study 
are as follows. We demonstrated that a population of 
long chains can be sustained by mutual catalysis 
sustained exclusively by template-assisted ligation.
This state is separated from the monomer-dominated one  
by a hysteretic first order phase transition 
(Eqs. (\ref{eq_c_down},\ref{eq_c_up}))
as a function of the concentration. 
We also demonstrated that the template-assisted ligation 
in our system is dominated by contributions from 
template-substrate pairs complementary over a well-defined length $k_0$, 
that is kinetically limited.
The average length of heteropolymers exponentially increases with 
$k_0$, with the upper bound given by a very simple expression, 
Eq. (\ref{eq_Lmax}), depending only 
on the ratio between the ligation and the breakage rates.

{\color{red} The spontaneous emergence of long polymers }
demonstrated in our study is of 
conceptual importance to 
the long-standing problem of the origin of life. 
Indeed, we offer a physically plausible path leading from the 
primordial soup dominated by monomers to a population of 
sufficiently long self-replicating chains.
This transition is one of the least understood 
processes in the RNA-world hypothesis.
%
%
%
%
%
It is known that functional RNA-based enzymes 
(ribozymes) need to be sufficiently long, which makes 
their spontaneous formation prohibitively unlikely.
According to our analysis, both the characteristic chain 
length and the minimal monomer concentration required 
for autocatalysis depend on the ratio of ligation and breakage rates. 
Large values of this ratio 
$\lambda/\beta \gg 1$ would allow long chains to form 
at physically possible concentrations $C \ll 1$M. One of the reasons that 
such spontaneous emergence of long-chained polymers 
has never been observed is that in experimental systems studied so far the ratio 
$\lambda/\beta$ remained low due to a very slow 
ligation process\cite{Szostak_1996}. Note that 
ligation and breakage processes in our system are not direct opposites 
of each other. Indeed, the ligation of e.g. nucleic acids requires 
activated terminal bases carrying free energy sufficient to 
form a new intra-polymer bond. To achieve the conditions necessary for
our autocatalytic regime one needs to either use heteropolymers 
chemically different from modern nucleic acids or to develop new 
activation pathways different from what has been used in experiments 
so far. The ligation can be further assisted e.g. by the 
absorption of polymers onto properly selected crystalline interfaces.
%

The present study was limited to the simplest version 
of the problem in which sequences of all heteropolymers 
were assumed to be completely random.  {\color{red} It provides a 
useful analytically solvable null-model against which future 
variants can be benchmarked.  Even though the informational entropy of 
the pool of polymers in our model is at its maximal value, the template-assisted ligation 
provides a mechanism for faithful transmission of information 
to the next generation.
We demonstrated that the spontaneous 
emergence of long chains is possible even 
in the limit where direct (non-templated) bond 
formation is negligible. This is especially important 
since non-templated polymerization is a regular 
equilibrium phenomenon and as such has a short memory.
In contrast, heritable transmission of sequence 
information via template-assisted ligation opens 
up an exciting possibility of long-term memory effects 
and ultimately of the Darwinian evolution 
in the space of polymer sequences.           
Incorporation of sequence effects is the logical 
next step in the development of our model, and we are currently working on it. }
There are several conceptually distinct yet non mutually exclusive 
scenarios giving rise to over-representation of certain sequences 
in the pool of heteropolymers. The first one is driven by the 
sequence dependence of model parameters such as hybridization free energies, 
fragmentation and ligation rates, and monomer composition of the primordial soup. 
The other scenario  is the spontaneous symmetry breaking in the sequence 
space\cite{Anderson,Goldenfeld}. Specifically, our results obtained within the 
Random Sequence Approximation need to be checked for local and global stability.
The local stability analysis deals with small deviations from a state in which 
populations of all sequences are equal to each other, while the global one
perturbs the system by strongly over-representing a small 
subset of sequences. {\color{red} This can be interpreted, correspondingly, 
as weak and strong selection limits.} Evidence of local or global 
instability would signal a symmetry breaking 
and would provide a scenario for the dramatic decrease 
in informational entropy of the population of polymers.
{\color{red} This is analogous to replica symmetry breaking 
suggested by P.W. Anderson \cite{Anderson} leading to a population 
dominated by a relatively small subset of mutually catalyzing sequences.}

\section*{Acknowledgements}
This research used resources of the Center for Functional Nanomaterials, which is a U.S. DOE Office of Science User Facility, at Brookhaven National Laboratory under Contract No. DE-SC0012704. Work at Biosciences 
Department was supported by US Department of
Energy, Office of Biological Research, Grant PM-031. 
We would like to thank Prof. Mark Lukin, Stony Brook University 
for valuable discussions.
%

\section*{Appendix A: $k$-mers and their hybridization dynamics}
To describe the hybridization dynamics during the night phase we 
introduce the concept of a $k-$mer defined as the segment of $k$ monomers 
with the specific sequence $\sigma$ within a longer chain of length $l \geq k$. 
Let $C \cdot p_{k}^{\left(\sigma\right)}$, 
be the concentration of $k-$mers with particular sequence $\sigma$. 
Let $C \cdot P_{k}$ be the concentration of all
$k-$mers of length $k$, regardless of their sequences. By definition,
$P_{k}=\sum_{\sigma}p_{k}^{\left(  \sigma\right)}$.
If all the sequences are completely random,
$p_{k}^{\left(  \sigma\right)  }=P_{k}z^{-k}$. 
Each chain of length $l$ contains
$\left( l+1-k\right)  $ '$k$-mers', therefore
\begin{equation}
P_{k}=\sum\limits_{l=k}^{\infty}\left(l+1-k\right)  f_{l}
\label{S1}
\end{equation}
Note that $P_{k}$ has the maximum value of $1$ which is approached in the limit
when all chains are much longer than $k$.

We consider a problem of hybridization of polymers 
since the start of the night phase of the cycle when all 
of them are not hybridized.
To describe the hybridization kinetics we use the fractions of fully
hybridized k-mers 
$1 \geq \varphi_{k}^{\left(  \sigma\right)  }\left(  t\right) \geq 0$
as our dynamic variables. By definition, the concentration of such pairs of
bound k-mers is $C\cdot p_{k}^{\left(  \sigma\right)  }\varphi
_{k}^{\left(  \sigma\right)  }\left(  t\right)  $. 
We note that hybridization states of different $k$-mers 
are not independent from each other since some of them overlap. 
To account for this, we introduce one more variable 
$\psi_{k}^{\left(\sigma\right)} \leq 1-\varphi_{k}^{\left(\sigma\right)}$ 
which is the fraction of all $k$-mers with a given sequence $\sigma$
that are available for hybridization. Now the binding kinetics of all $k-$mers
can be described by the following set of coupled kinetic equations:%
\begin{equation}
\tau \dot{\varphi}_{k}^{\left(  \sigma\right)  }=C\cdot p_{k}^{\left(
\sigma'\right)  }\psi_{k}^{\left(  \sigma'\right)  }\psi
_{k}^{\left(  \sigma\right)  }-\exp\left(  \frac{\Delta G_{\sigma}%
}{k_{B}T}\right)  \varphi_{k}^{\left(  \sigma\right)  } \label{kin}%
\end{equation}
Here $1/\tau$ is the hybridization rate, $\Delta G_{\sigma}$ is
the hybridization free energy for a given sequence $\sigma$, 
and $ \sigma'$ is the sequence complementary to 
$\sigma$. For simplicity, we 
consider a symmetric case where mutually complementary $k$-mers have the same
fraction, $p_{k}^{\left(  \sigma\right)  }=p_{k}^{\left(  \sigma'\right)
}$ . In order to solve these equation, one needs to specify a relationship
between fraction of available k-mers $\psi_{k}^{\left(  \sigma\right)  }$ and
hybridization probabilities, $\varphi_{k}^{\left(  \sigma \right)  }$, that 
would take into account mutual overlap of the sequences.
However, at early stages the hybridization probability remains sufficiently
low, and one can therefore assume $\psi_{k}^{\left(  \sigma\right)  }=\psi
_{k}^{\left(  \sigma'\right)  }\approx 1$ in Eq.  (\ref{kin}). 
This results in a set of decoupled equations 

\begin{equation}
\tau\dot{\varphi}_{k}^{\left(  \sigma\right)  }=C\cdot p_{k}^{\left(
\sigma\right)  }-\exp\left(  \frac{\Delta G_{\sigma}}{k_{B}T}\right)  \varphi_{k}^{\left(  \sigma\right)  } 
\end{equation}
The solution is the exponential relaxation of hybridization
variables $\varphi_{k}^{\left(  \sigma\right)  }$ 
towards their equilibrium values:
\begin{equation}
\varphi_{k}^{\left(  \sigma\right)  }\left(  t\right)  =K_{k}^{\left(
\sigma\right)  }p_{k}^{\left(  \sigma\right)  }\left(  1-\exp\left(
-\frac{t}{\tau_{k}^{\left(  \sigma\right) }}\right)  \right)
\end{equation}
In this expression
\begin{equation}
K_{k}^{\left(  \sigma\right)  }=C\exp\left(  -\frac{\Delta
G_{\sigma}}{k_{B}T}\right)
\end{equation}%
\begin{equation}
\tau_{k}^{\left(  \sigma\right)  }=\tau\exp\left( - \frac{\Delta
G_{\sigma}}{k_{B}T}\right) \qquad .
\end{equation}

The single most important factor that determines the hybridization free
energy\ $\Delta G_{\sigma}$ is the sequence length $k$. For simplicity of the
analysis we will replace $K_{k}^{\left(  \sigma\right)  }$ with its sequence-
averaged value:%
\begin{equation}
K_{k}^{\left(  \sigma\right)  }\approx K_{k}=C\exp\left(
-\frac{\Delta G_{0}+k \Delta G}{k_{B}T}\right)
\end{equation}
This leads to the following result:%
\begin{eqnarray}
\varphi_{k}\left(  t\right)  =C P_{k}z^{-k}\exp\left(  -
\frac{\Delta G_{0}+k \Delta G}{k_{B}T}\right)  \cdot  \nonumber \\
\cdot \left(  1-\exp\left[ 
-\frac{t}{\tau}\exp\left(\frac{\Delta G_{0}+k \Delta G}{k_{B}T}\right) \right]  \right)
\label{S8}
\end{eqnarray}

As shown in Figure 4 at any given time $t$ this expression is strongly
peaked at a single value of $k$, which weakly (logarithmically) 
depends  on time:%
\begin{equation}
k\approx k_{0}\left(  t\right)  \simeq -\frac{k_{B}T}{\Delta G}\log\left( \frac{t}{\tau}\right)
\label{S9}
\end{equation}%
\begin{equation}
\varphi_{k_{0}}\simeq CP_{k_{0}}\exp\left(  -\frac{\Delta G_{0}}%
{k_{B}T}+\varepsilon k_{0}\right)
\end{equation}
\section*{Appendix B: Evaluating the effects of a finite $\bar{L}$. }
The equations (\ref{eq_fl}) and (\ref{eq_sc_SI}) in the main text were derived in the 
limit $\bar{L} \gg  1$. Below we will relax these approximations
to derive the exact formula working for arbitrary $\bar{L}$.

In deriving the Eq. (\ref{eq_fl}) in the main text we 
replaced the discrete summation with an integral. 
This approximation can be avoided by performing 
an explicit summation of the discrete geometric progression:
\begin{equation}
\sum_{l=1}^{\infty} l \cdot \exp(-\frac{l}{\bar{L}})=
\frac{\exp\left(-\frac{1}{\bar{L}}\right)}
{\left[1-\exp\left(-\frac{1}{\bar{L}}\right)\right]^2}=
\frac{1}{4\sinh\left(\frac{1}{2\bar{L}}\right)^2}  \quad .
\end{equation}

This amounts to replacing $\bar{L}$ in Eq. (3) with $\frac{1}{2\sinh\left(\frac{1}{2\bar{L}}\right)}$:
\begin{equation}
\frac{1}{2\sinh\left(\frac{1}{2\bar{L}}\right)}=\sqrt{\frac{\mu}{\beta}} \qquad .
\label{SX}
\end{equation}

The exact triple summation of the Eq. (\ref{eq_lambda}) in the main text
\begin{eqnarray}
\mu=\lambda \left(\frac{C}{C_0}\right)^2
\sum_{k_1=1}^{k_0}\exp(k_1 \cdot \epsilon) \sum_{k_2=1}^{k_0}\exp(k_2 \cdot \epsilon) \cdot  \nonumber \\
\cdot \sum_{l=k_1+k_1}^{\infty}(l-k_1-k_2+1)f_l \quad . 
\end{eqnarray}
for $f_l \sim \exp(-l/\bar{L})$ can be carried out in two steps.
First, the sum over $l$ combined with normalization $\sum_{l} l \cdot f_l=1$
gives rise to 
\begin{eqnarray}
\mu=\lambda \left(\frac{C}{C_0}\right)^2 \exp(1/\bar{L})
\sum_{k_1=1}^{k_0}\exp \left[k_1 \cdot  (\epsilon-1/\bar{L})\right] \cdot \nonumber \\
\cdot \sum_{k_2=1}^{k_0}\exp \left[k_2 \cdot (\epsilon-1/\bar{L}) \right] \quad .
\end{eqnarray}
The discrete summation over $k_1$ and $k_2$ results in
\begin{equation}
\mu=\lambda \left(\frac{C}{C_0}\right)^2 \exp(1/\bar{L})
 \left( \frac{\exp[k_0(\epsilon -1/\bar{L})] -1}{1-\exp(-\epsilon+1/\bar{L})}\right)^2 \qquad .
\label{eq_mu1}
\end{equation}
The Eq. (\ref{eq_sc_SI}) then becomes
\begin{eqnarray}
\frac{1}{2\sinh\left(\frac{1}{2\bar{L}}\right)}\exp\left(\frac{k_0-1/2}{\bar{L}}\right)= \nonumber \\
=\frac{C}{C_0} \cdot \sqrt{\frac{\lambda}{\beta}} 
\cdot \frac{\exp\left(k_0 \epsilon \right)-\exp\left(k_0/\bar{L} \right)}{1-\exp(-\epsilon+1/\bar{L})} \quad .
\label{eq_sc}
\end{eqnarray}
Here we neglected the exponentially small term in the enumerator of the r.h.s. of 
Eq. (\ref{eq_mu1}).
The dashed line in Fig. 4 shows $\bar{L}$ defined by this equation 
plotted as a function of $C$.

\section*{Appendix C: Ligation-fragmentation kinetics}
The Eq. (\ref{eq_sc_SI}) describes the effective merger rate $\mu$
when lengths $n$ and $m$ of two substrate chains hybridized 
to a template are longer than $k_{0}$. In a more general case 
one needs to introduce length-dependent effective merger rate
$\mu_{nm}$.
Under RSA this rate is given by:
\begin{eqnarray}
&&\mu_{nm}=\lambda C^2\sum\limits_{k_{1}=1}^{\min\left(
n,k_{0}\right)  }\sum\limits_{k_{2}=1}^{\min\left(  m,k_{0}\right)  }%
\frac{P_{k_{1}+k_{2}}}{z^{k_{1}+k_{2}}} \cdot \nonumber \\
&&\cdot \exp\left(  -\frac{2\Delta G_{0}+\left(k_{1}+k_{2}\right) \cdot \Delta G}{k_{B}T}  \right) = \nonumber \\
&&=\lambda \left (\frac{C}{C_0}\right)^2 \cdot \sum\limits_{k_{1}=1}^{\min\left(
n,k_{0}\right)  } 
\sum\limits_{k_{2}=1}^{\min\left(  m,k_{0}\right)  }%
P_{k_{1}+k_{2}} \cdot \nonumber \\
&& \cdot \exp \left( \left(  k_{1}+k_{2}\right)\cdot \epsilon  \right)
\end{eqnarray}
Here $\mu_{nm}$ corresponds to a particular order in which 
chains $n$ and $m$ merge into a longer chain. 
Note that for directed chains such as 
nucleic acids there are two ways of merging chains, while 
for undirected polymers there are four. 
 
For nucleic acids, when 
two chain segments are bound to the same template and are 
directly adjacent to each other (Fig. 1ab) there is an additional gain 
in free energy $\Delta G_{st}$ due to stacking. 
It is straightforward to incorporate $\Delta G_{st}$ into our formalism
by redefining $C_{0}$ as $C_0=\exp[-(\Delta G_0+\Delta G_{st}/2)/k_{B}T]$ 
(in molar).

For directed polymers the resulting set of kinetic equations can be written as:%
\begin{eqnarray}
&&\frac{1}{2\beta}\dot{f}_{n} = -\left[  \frac{n}{2}+\Gamma^{2}\sum\limits_{m}%
\mu_{n,m}f_{m}\right]  f_{n}+\sum\limits_{m>n} f_{m}+\nonumber \\
&&+\Gamma^{2}\sum\limits_{m<n}\frac{\left(
1+\delta_{n-m,m}\right)  }{2}\mu_{m,n-m}f_{m}f_{n-m} \qquad . 
\label{kin2_SI}%
\end{eqnarray}
Here $\Gamma$ is the dimensionless control parameter of the model which is
proportional to monomer density:%
\begin{equation}
\Gamma=\left(\frac{C}{C_{0}}\right)\sqrt{\frac{\lambda}{\beta}} \label{nu_SI}%
\end{equation}
and $\mu_{nm}$ is the "$k$-mer"- dependent ligation matrix:%
\begin{equation}
\mu_{nm}=\sum\limits_{k_{1}=1}^{\min\left(  n,k_{0}\right)  }%
\sum\limits_{k_{2}=1}^{\min\left(  m,k_{0}\right)  }P_{k_{1}+k_{2}}\exp\left(
\epsilon \cdot \left(  k_{1}+k_{2}\right)  \right) \qquad .
\label{S17}
\end{equation}
This set of kinetic equations gives a complete description of 
the system in question and was numerically integrated to compare 
with our analytical results.



\begin{thebibliography}{99}
\bibitem{Chaikin} T. Wang, R. Sha, R. Dreyfus, M. E. Leunissen, 
C. Maass, D. J. Pine, P. M. Chaikin, and N. C. Seeman, 
Nature {\bf 478}, 225 (2011).
\bibitem{Brenner} Z. Zeravcic, and M. P. Brenner,  
Proc. Natl. Acad. Sci. USA {\bf 111}, 1748 (2014).
\bibitem{Eigen} M. Eigen and P. Schuster, 
Naturwissenschaften {\bf 64}, 541 (1977).
\bibitem{Dyson} F. J. Dyson, 
Journal of Molecular Evolution {\bf 18}, 344 (1982).
\bibitem{Kauffman} S.A. Kauffman,  
Journal of Theoretical Biology {\bf 119}, 1 (1986).
\bibitem{Jain} S. Jain and S. Krishna, 
Phys. Rev. Lett. {\bf 81}, 5684 (1998).
\bibitem{Szostak_1989} J. A. Doudna, J. W. Szostak, 
Nature {\bf 339}, 519 (1989). 
\bibitem{Joyce} T. A. Lincoln and G. F. Joyce,  
Science {\bf  323}, 1229 (2009).
\bibitem{Gilbert} W. Gilbert, 
Nature {\bf 319}, 618 (1986).
\bibitem{Orgel} L. E. Orgel, 
Critical Reviews in Biochemistry and Molecular Biology {\bf 39}, 99 (2004).
\bibitem{Joyce_Review} M. P. Robertson and G. F. Joyce, 
Cold Spring Harbor Perspectives in Biology
{\bf 4}, a003608 (2012).
\bibitem{Braun} C. B. Mast, S. Schink, U. Gerland, and D. Braun,
Proc. Natl. Acad. Sci. USA {\bf 110}, 8030 (2013).
\bibitem{Anderson} P.W. Anderson, 
Proc. Natl. Acad. Sci. USA {\bf 80}, 3386 (1983).
\bibitem{Anderson_Stein}
P. W. Anderson, and D. L. Stein, in {\it Self-Organizing Systems} edited by  
F. E. Yates, A. Garfinkel, D. O. Walter, and G. B. Yates, (Springer US, 1987) 
pp. 445-457.
\bibitem{Derr} 
J. Derr, M. L. Manapat, S. Rajamani, K. Leu, R. Xulvi-Brunet, 
I. Joseph, M. A. Nowak, and I. A. Chen, 
Nucleic Acids Research {\bf  40}, 4711 (2012).
\bibitem{Bellini} 
T. Bellini, G. Zanchetta, T. P. Fraccia, R. Cerbino, E. Tsai, 
G. P. Smith, M. J. Moran, D. M. Walba, and N. A. Clark,  
Proc. Natl. Acad. Sci. USA {\bf  109}, 1110 (2012).
\bibitem{Szostak_1996} R. Rohatgi, D.P. Bartel, J. W. Szostak, 
J. Am. Chem. Soc. {\bf 118}, 3332 (1996).
\bibitem{Goldenfeld} K. Vetsigian and N. Goldenfeld,  
Proc. Natl. Acad. Sci. USA {\bf 106}, 215 (2009).
\end{thebibliography}
\end{document}